\documentclass[runningheads]{llncs}
\usepackage[T1]{fontenc}
\usepackage{graphicx}
\usepackage{amsmath}
\usepackage{bm}
\usepackage{esvect}
\usepackage[version=4]{mhchem}
\usepackage[utf8]{inputenc} 
\usepackage{mathtools}

\usepackage[table]{xcolor}% http://ctan.org/pkg/xcolor
\graphicspath{./images/}

\usepackage[numbers]{natbib}

\begin{document}
\title{Plumbing Analog of Molecular Computation}
%
%\titlerunning{Abbreviated paper title}
% If the paper title is too long for the running head, you can set
% an abbreviated paper title here
%
\author{Roger D. Jones\inst{1,2}\orcidID{0000-0002-8491-5421} \and
Achille Giacometti\inst{1}\orcidID{0000-0002-1245-9842} \and
Alan M. Jones\inst{2,3}\orcidID{0000-0002-2365-6462}}
\authorrunning{R. Jones, A. Giacometti, A. Jones}
% First names are abbreviated in the running head.
% If there are more than two authors, 'et al.' is used.
%
\institute{European Centre for Living Technology,  
Department of Molecular Sciences and Nanosystems,
Ca’ Foscari, University of Venice, Italy \and
Department of Biology, University of North Carolina at Chapel Hill, USA 
\and
Department of Pharmacology, University of North Carolina at Chapel Hill, USA 
\email{RogerDJonesPhD@gmail.com}\\
}
\maketitle              % typeset the header of the contribution
\begin{abstract}
Biological information processing often arises from mesoscopic molecular systems operating far from equilibrium, yet their complexity can make the underlying principles difficult to visualize. In this study, we introduce a macroscopic hydraulic model that serves as an intuitive analog for the molecular switching behavior exhibited by G protein–coupled receptors (GPCRs) on the cell membrane. The hydraulic system reproduces the essential structural and functional features of the molecular switch, including the presence of up to three distinct steady-state solutions, the characteristic shapes of these solutions, and the physical interpretation of the control parameters governing the behavior of the system. By mapping water flow, energy barrier height, and siphoning dynamics onto biochemical flux, activation energy, and state transitions, the model provides a transparent representation of the mechanisms that regulate GPCR activation.

The correspondence between the hydraulic analog and the molecular system suggests several experimentally testable hypotheses about GPCR function. In particular, the model highlights the central role of energy flux—driven by imbalances in ATP/ADP or GTP/GDP concentrations, in activating the molecular switch and maintaining nonequilibrium signaling states. It also identifies two key parameters that primarily determine switch behavior: the energy difference between the active and inactive states and the effective height of the energy barrier that separates them. These results imply that GPCR signaling dynamics may be governed by generalizable physical principles rather than by biochemical details alone. The hydraulic framework thus offers a tractable platform for interpreting complex molecular behavior and may aid in the development of predictive models of GPCR function in diverse physiological contexts.

\end{abstract}
\section{Introduction}

\begin{figure}
\includegraphics[width=\textwidth]{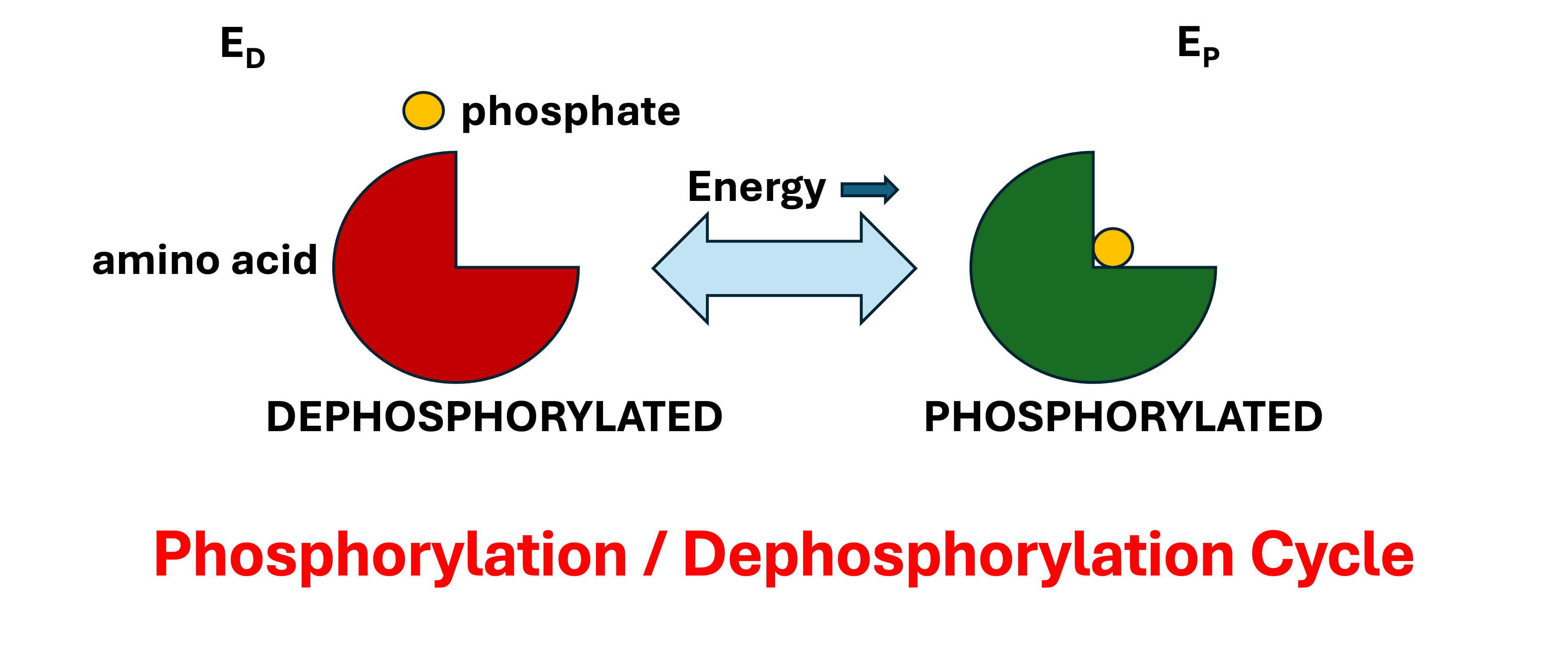}
\caption{A phosphorylation–dephosphorylation cycle used as a biological switch. Energy input, typically through ATP hydrolysis, raises the dephosphorylated state of energy $E_d$ to the phosphorylated state of energy $E_p$. The protein matrix, in which the switch is embedded, modulates these energies through conformational changes, and the phosphorylation state determines the downstream cellular response.} \label{PdPC}
\end{figure}

Life is an organized, information-processing phenomenon inseparably embedded in a turbulent physical environment. The persistence of life requires continuous adaptation, and adaptation, in turn, depends on the acquisition, transmission, and transformation of information. In this sense, biological function is a form of computation. At the molecular scale, this computation is executed by regulatory processes that operate analogously to switches \cite{jones14model,jones2023proposed,jones2024information}. These mechanisms, driven by external energy sources, such as sunlight or nutrient-derived metabolites, control cellular behavior through tightly coordinated biochemical reactions. The phosphorylation–dephosphorylation cycle (PdPC) (Figure \ref{PdPC}) exemplifies such a switch \cite{qian2007phosphorylation}. Other molecular switches, including GTPases, differ in biochemical detail but share the same fundamental operational principles. 
Here, we use the PdPC as a stand-in for all types of biological switch that can be characterized by a simple chemical reaction like the one illustrated in Figure \ref{PdPC}.

These logically reversible reactions, which usually target amino acids within proteins, create decision points that guide cellular responses in diverse physiological settings. Through the collective dynamics of many of such switches, cells regulate metabolism, immunity, and behavior, allowing adaptive biological computation to emerge from the continuous reconfiguration of molecular states.

\begin{figure}[hbt!] %s state preferences regarding figure placement here
% use to correct figure counter if necessary
%\renewcommand{\thefigure}{2}
\includegraphics[width=1.0\textwidth]{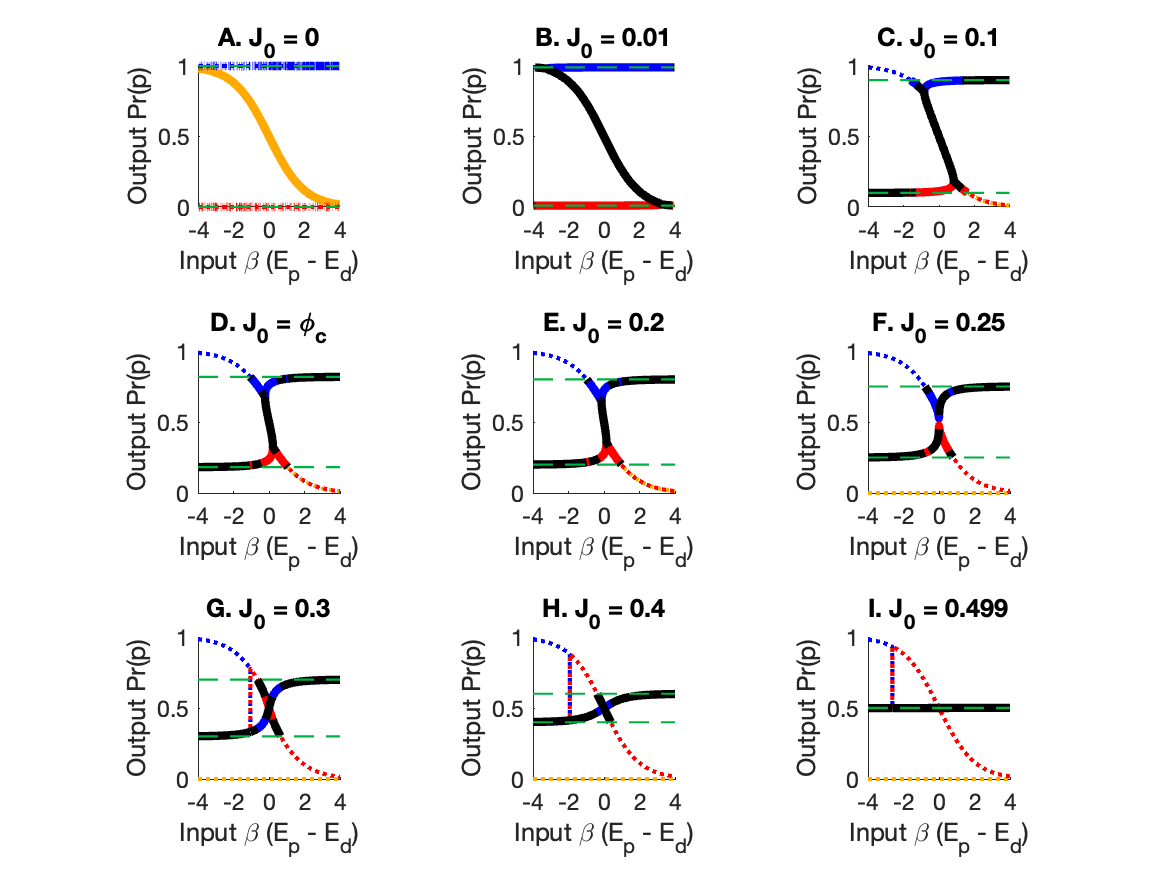}
\caption{
Information-maximizing dynamical paths (black). A: In the zero-flux limit, all paths transmit equivalent information. B: With a small nonzero flux $J_0$, the thermodynamic branch (yellow in A) attains the highest capacity, and the optimal path for $Pr(p)$, the probability of finding the switch phosphorylated, follows the canonical ensemble. B–I: For fixed flux $J_0$, changes in the binding-energy difference $\beta(E_p-E_d)$ generate discontinuities in the high-capacity path. The global maximum corresponds to the dominant quasistable state; other paths represent local maxima.
}
\label{Solutions} % \label works only AFTER \caption within figure environment
\end{figure}

Simple molecular switches of this kind can be analyzed using statistical-physics approaches \cite{jones2025information} similar to those outlined by Seifert \cite{seifert2008stochastic}. Assuming that natural selection favors efficient information processing, switches can occupy quasistable on, off, and intermediate configurations \cite{jones2025information}. These states are shown in Fig. \ref{Solutions}.

The architecture of the on, off, and intermediate states of a switch is governed by two external control parameters: the chemical flux $J_0$ passing through the switch and the energy difference $\beta(E_p-E_d)$ between the phosphorylated and dephosphorylated states, where $\beta$ is the inverse temperature of the surrounding heat bath. 
Additionally, the transitions among the states are facilitated by activation energy barriers between states.

In Fig. \ref{Solutions}A, which illustrates solutions with no external chemical flux, the yellow curve represents the traditional thermodynamic equilibrium solution, while the red and blue curves denote additional branches that appear for switches. 
The remaining figures show modifications to the equilibrium solutions as the chemical flux $J_0$ increases.
Black overlays highlight information-efficient solutions, and dotted curves correspond to nonphysical solutions that violate the conditions for probability. 
Under optimal-efficiency assumptions, molecular switches admit not merely two discrete states but as many as three distinct quasistates.

\begin{figure}[hbt!] %s state preferences regarding figure placement here
% use to correct figure counter if necessary
%\renewcommand{\thefigure}{2}
\centering
\includegraphics[width=0.7\textwidth]{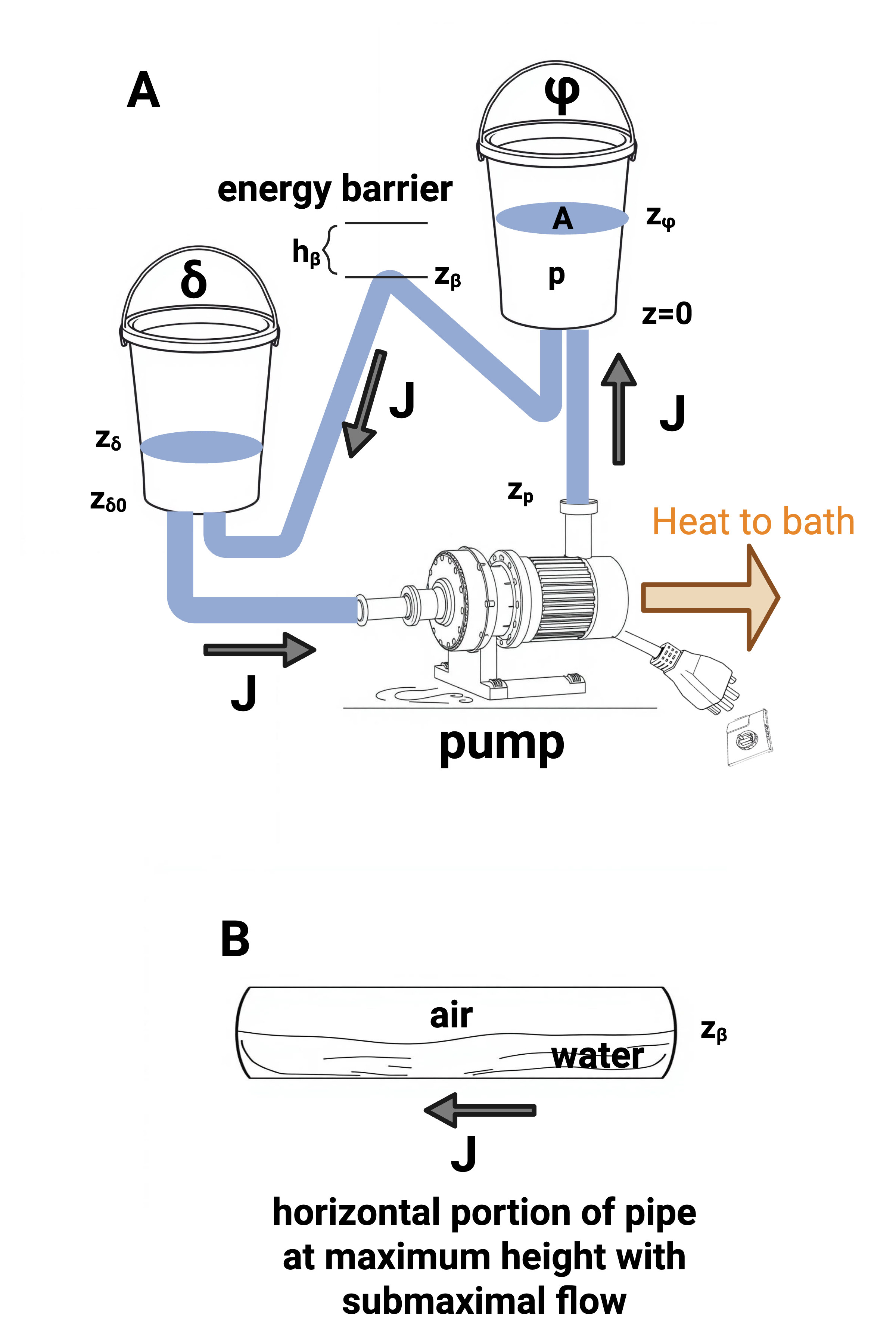}
\caption{ 
Macroscopic hydraulic analogy of a molecular switch. {\bf A.} Siphoning regime: Water levels in the $\varphi$ and $\delta$ buckets represent the probabilities of the switch being phosphorylated ($p$) or dephosphorylated. 
A powered pump supplies the driving flux, while the height of the return tube acts as an energy barrier that regulates reverse flow. 
{\bf B.} 
Non-siphoning regime: At specific barrier heights $z_{\beta}$, the filling history allows air to enter the pipe at the barrier, preventing siphoning. 
In this state, a current $J$ may still flow, but the system does not siphon, producing alternative, non-siphoning solutions shown in Figure \ref{WaterExperiment3}.
}
\label{WaterExperiment} % \label works only AFTER \caption within figure environment
\end{figure}

Because the resulting behavior is more complex than that of engineered two-state switches, a simple macroscopic analogy is useful for conveying the underlying principles. The essential features of Fig. \ref{Solutions}, an externally supplied flux that powers information processing, an energy barrier that modulates state occupancy, and thermal noise that introduces stochasticity, can be captured by a closed hydraulic system. Figure \ref{WaterExperiment} depicts such a model: water circulates through pipes and buckets, driven by a pump that supplies the flux $J$; the height of a return pipe represents the activation energy. 
We will see that the characteristic length scale determines the effective temperature of an effective background bath.
The heat  lost from the pump dissipates into the surrounding environment, analogously to that in the biological heat bath.

Because the hydraulic model incorporates the essential physical features of the molecular switch, driving flux, an adjustable energy barrier, and thermally induced fluctuations, it reproduces the qualitative behavior of statistical-mechanical solutions in Fig. \ref{Solutions}. In this study, we analyze the dynamics of the model shown in Fig. \ref{WaterExperiment}. We test the theoretical predictions experimentally in follow-up work.

\begin{figure}[hbt!] %s state preferences regarding figure placement here
% use to correct figure counter if necessary
%\renewcommand{\thefigure}{2}
\centering
\includegraphics[width=1.0\textwidth]{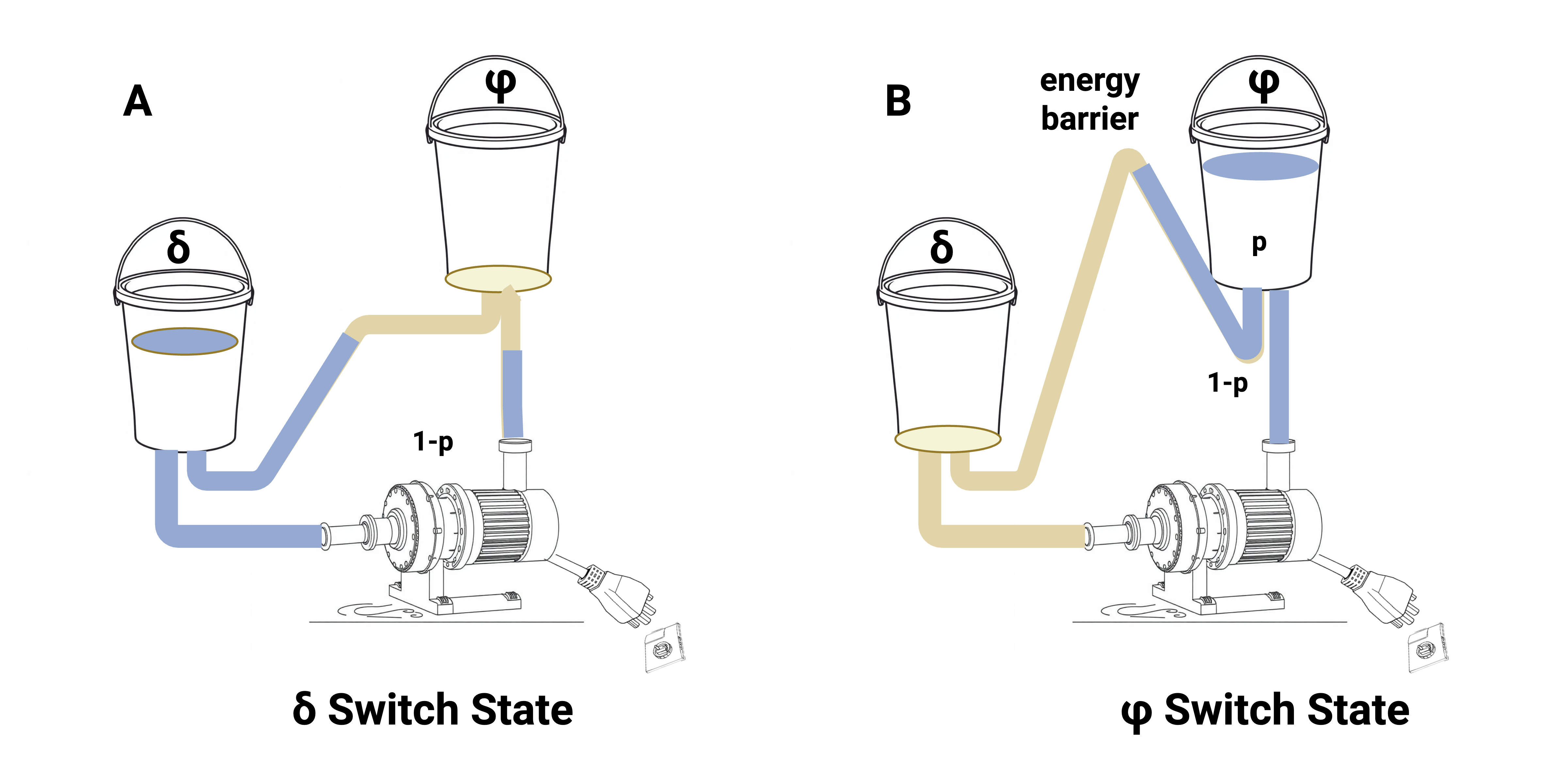}
\caption{ 
Non-siphoning solutions. {\bf A.} A non-siphoning state in which the switch is predominantly dephosphorylated. {\bf B.} A complementary state in which the switch is predominantly phosphorylated. For specific barrier heights, a small current $J$ may still flow, but otherwise the system remains stationary with no net flow.
}
\label{WaterExperiment3} % \label works only AFTER \caption within figure environment
\end{figure}

The conceptual framework developed here also informs our understanding of one of the most significant biological information-processing nanosystems: G protein–coupled receptors (GPCRs). Their relevance and connection to this work are explored further in the Discussion section.

\section{Methods}

The first step is to determine the probability of phosphorylation 
$p$ as a function of the system parameters defined in Fig. \ref{WaterExperiment}. These parameters include the water volumes in the two buckets, their cylindrical cross-sectional areas, the diameters and cross-sectional areas of the connecting pipes, the pump-generated current 
$J$, the head 
$h$, and the water heights throughout the pipes and buckets.

We compare the relations of the macroscopic systems in Figures \ref{WaterExperiment} and \ref{WaterExperiment3}
with the mesoscopic biological system shown in Figure \ref{Solutions}
to gain insight into the mechanisms of molecular switching.

\subsection{Three Solutions}

The three solutions for the water flow in the system are shown in Figures \ref{WaterExperiment} and \ref{WaterExperiment3}.
The system is composed of two buckets labeled $\delta$ and $\varphi$, dephosphorylation and phosphorylation, respectively.
A pump drives water through the system, and pipes connect the buckets and the pump.

The first solution is the siphoning solution in which the water flows continuously to both buckets $\varphi$ and $\delta$ in a cycle.
There are two other solutions in which the water is concentrated in the $\varphi$ bucket or in the $\delta$ bucket.

\subsubsection{Siphon solution}
The probability $p_s$ of finding a water molecule in the $\varphi$ bucket is

\begin{equation} \label{siphonProbability}
    p_s=\frac{V_{\varphi}}{V}
    =
    \frac{A\;z_{\phi}}{V}
    =\frac{z_{\varphi}}{L_c} \coloneqq  \hat z _{\varphi}
\end{equation}
where 
$V_{\varphi}$ is the volume of water in the phosphorylation bucket,
$V$ is the total volume of water,
$A$ is the cross-sectional area of the phosphorylation bucket (assumed cylindrical),
$z_{\varphi}$ is the height of water in the phosphorylation bucket,
$L_c$ is a characteristic length scale,
\begin{equation}\label{CharLength}
    L_c \coloneqq \frac{V}{A}
\end{equation}
and $\hat z_{\varphi}$ is a normalized value of the height of the water.

The head loss $h_{\beta}$ is calculated with the Darcy-Weisbach equation \cite{white1999fluid}: 
\begin{equation}
    h_{\beta}=
    z_{\varphi} -z_{\beta}
    = f \frac{l_{\varphi} v^2}{2 g d}
\end{equation}
where $f$ is the Darcy friction factor, $l_{\varphi}$ is the length of pipe from $z=0$ to $z_{\beta}$, $v$ is the average speed of the water in the pipe, $g$ is the gravitational constant, and $d$ is the diameter of the pipe.

The flow per water of a water molecule is

\begin{equation}
     J =   \frac{ a v}{V}   
\end{equation}
where
\begin{equation}
    a = \pi \left(  \frac{d}{2}   \right)^2
\end{equation}
is the cross-sectional area of the pipe.

\begin{equation}
    v=
    \frac{V J}{a}
    =
    \frac{L_c J}{\hat a} 
\end{equation}
where
\begin{equation}
    \hat a \coloneqq \frac{a}{A}
\end{equation}

The expression for the head loss becomes
\begin{equation}
    \hat h_{\beta}=
    \hat z_{\varphi} - \hat z_{\beta}
    = \hat J ^2
\end{equation}
or
\begin{equation} \label{zb}
    \hat z_{\beta} =
    \hat z_{\varphi} - \hat J^2
    =
    p_s - \hat J^2
\end{equation}
where
\begin{equation}\label{tcJ}
    \hat J^2 
    =
    (t_c J )^2
\end{equation}
and
\begin{equation}
    t_c^2 = \frac{f \hat l_{\phi} L_c^2}{2gd \hat a^2}
\end{equation}
The length expressions with hats are quantities normalized to the characteristic length $L_c$ or, in the case of the pipe area $a$, to the cross-sectional area of the $\varphi$ bucket.

\begin{equation}
    \hat h_{\delta}
        =
    \hat z_{\beta} - \hat z_{\delta} + \hat z_{\delta 0}
    = \left( \frac{l_{\delta }}{ l_{\phi}}  \right)   \hat J ^2
\end{equation}
or
\begin{equation}
    \hat z_{\beta}
    =\hat z_{\delta} - \hat z_{\delta 0} 
     + \left( \frac{l_{\delta }}{ l_{\phi}}  \right)    \hat J ^2
     =
     p_s
     -  \hat J ^2
     \;\;\;
     \mbox{from Eq. \ref{zb}}
\end{equation}
where $p_s=z_{\varphi}$ is the probability of phosphorylation in the siphoning solution and $l_{\delta}$ is the length of the pipe from $z_{\beta}$ to $z_{\delta 0}$.
\begin{equation}\label{siphoningSolution}
    p_s
    = \Delta \hat z
     + \left(1+ \frac{l_{\delta }}{ l_{\phi}}  \right)   \hat J ^2
     \;\;\;\; \forall\;
     0< z_{\beta}< p_s
\end{equation}
where
\begin{equation}\label{DeltaZ}
    \Delta \hat z =
    \hat z_{\delta} - \hat z_{\delta 0} 
\end{equation}
We can write $\Delta z$ in terms of energy as in the case of the molecular switch, the x-axis in Figure \ref{Solutions}.
\begin{equation}\label{betaa}
    \Delta z   =
    \beta_a (E_{\delta}-E_{\delta 0})
\end{equation}
where 
\begin{equation}
    E_{\delta} = m g z_{\delta}
\end{equation}

\begin{equation}
    E_{\delta 0}  = m g z_{\delta 0}
\end{equation}
$m$ is the mass of one mole of water,
and
\begin{equation}\
    \beta_a = \frac{A}{mg V}
    =
    \frac{1}{mg L_c}
\end{equation}
Comparing Equation \ref{betaa} with the x axes in Figure \ref{Solutions} indicates that $\beta_a$ is the analog of the inverse temperature $\beta$ in the mesoscopic model.
The energy difference of the potential $E_{\delta}-E_{\delta 0}$ corresponds to $E_{p}-E_{d }$ in the microscopic model.
Comparison of Equation \ref{tcJ} with the microscopic model indicates that $t_cJ$ is the analog of $J_0$.

\subsubsection{$\delta$ solution}
For the $\delta$ solution, the current $J$ is zero. 
\begin{equation}
    p_{\delta} =0
    \;\;\;\; \forall\;
     z_{\beta}  
\end{equation}

\subsubsection{$\varphi$ solution}
The probability $p_{\varphi}$ of phosphorylation in the $\delta$ state (Figure \ref{WaterExperiment3}A) is
\begin{equation}
    p_{\varphi} = \frac{V_{\varphi}}{V}=
    1- \frac{a}{A} \frac{l_w}{L_c}
    = 1-a_c \hat l_w
\end{equation}
where $l_w$ is the length of pipe that holds water in Figure \ref{WaterExperiment3}B.
\begin{equation}
    \hat l_w = z_{\varphi} + \hat z_p
    = p_{\varphi} + \hat z_p
\end{equation}
where $z_p$ is the height of the pump.
The probability $p_{\varphi}$ is then
\begin{equation}
    p_{\varphi} =
    \frac{1- \hat a \; \hat l_w}{1+\hat a}
    \;\;\;\;\forall\;  z_{\beta} \ge z_{\varphi}
\end{equation}

\section{Results}

\begin{figure}[hbt!] %s state preferences regarding figure placement here
% use to correct figure counter if necessary
%\renewcommand{\thefigure}{2}
\includegraphics[width=1.0\textwidth]{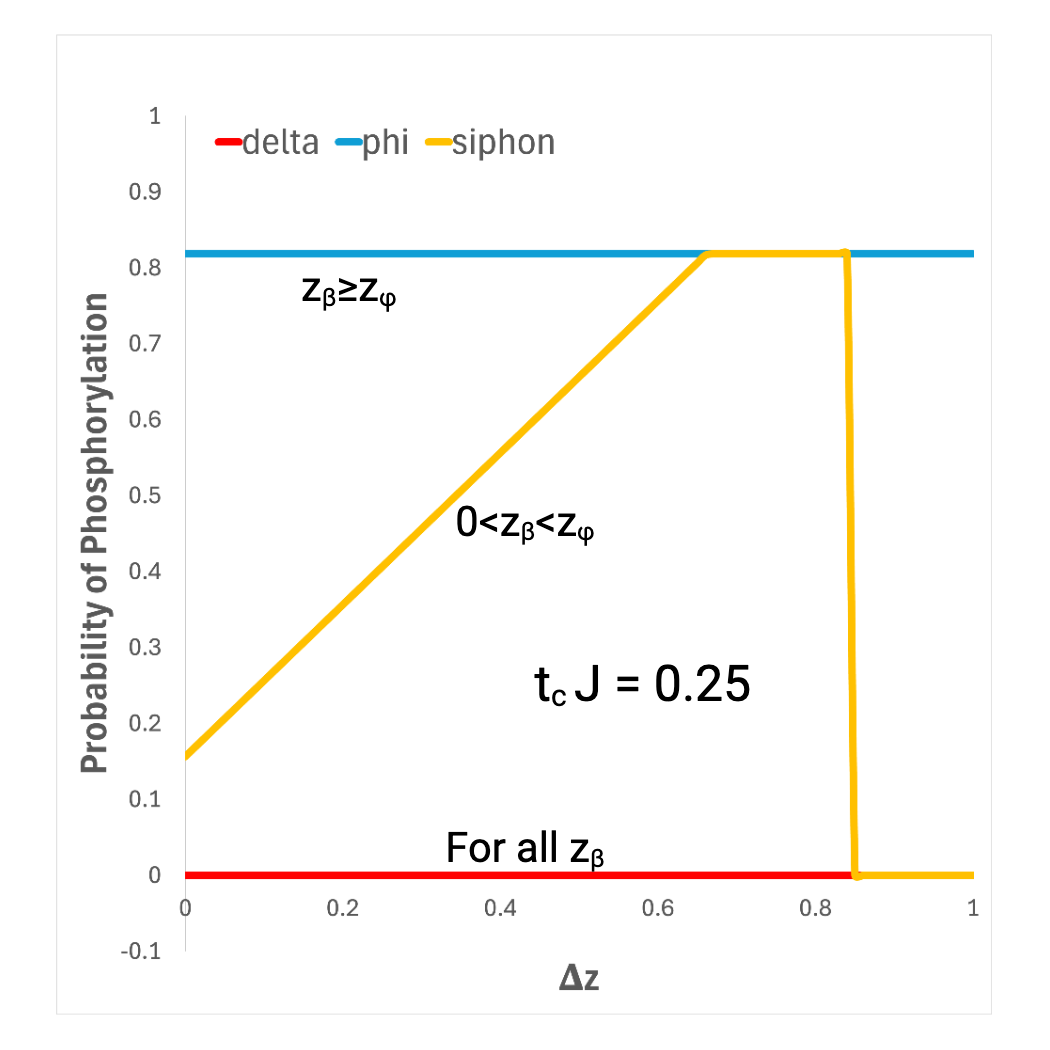}
\caption{
Example set of three solutions for the hydraulic model in Figs. \ref{WaterExperiment} and \ref{WaterExperiment3}. The blue curve represents the $\varphi$ solution, the red curve the $\delta$ solution, and the yellow curve the siphoning solution. When the barrier height 
$z_{\beta}$
 exceeds the maximum water height in the $\varphi$ bucket, only the $\varphi$ and $\delta$ solutions are possible. When 
$z_{\beta}$
 lies between the maximum $\varphi$ height and the bottom of the bucket, both the siphoning and $\delta$ solutions can occur. For still lower barrier heights, the $\delta$ solution is the only remaining physically realizable state.
}
\label{Plumbing} % \label works only AFTER \caption within figure environment
\end{figure}

When the barrier height 
$z_{\beta}$
 exceeds the maximum water height in the $\varphi$ bucket, only the $\varphi$ and $\delta$ solutions are possible. When 
$z_{\beta}$
 lies between the maximum $\varphi$ height and the bottom of the bucket, both the siphoning and $\delta$ solutions can occur. For still lower barrier heights, the $\delta$ solution is the only remaining physically realizable state.
 Note that the solutions for the macroscopic switch shown in Figure \ref{Plumbing} for $\hat J=t_c J=0.25$ have a similar form as the set of mesoscale solutions shown in Figure \ref{Solutions} for $J_0=0.25$.

The height of the energy barrier and the imposed water flow determine the state of the hydraulic switch for a fixed current 
$J$. For some values of the barrier height 
$z_{\beta}$, multiple solutions can coexist. When the barrier exceeds the maximum water height in the $ \varphi $ bucket, only the $\varphi$ solution is possible. 
As the barrier is lowered below this height, water begins to drain from the bucket and the system transitions to the siphoning solution. In the transition, the $\varphi$ and siphon solutions may coexist, corresponding to the flat regions that overlap the blue and yellow curves in Fig. \ref{Plumbing}. With further reduction of the barrier and in the absence of pumping, only the $\delta$ solution remains accessible.
The details of these transitions will be examined experimentally in a follow-up study.

\section{Discussion}

In this study, we introduced a macroscopic hydraulic system as an intuitive analog for the mesoscopic molecular information-processing machinery of switches on the cell membrane. The correspondence between the two systems is striking: both exhibit up to three distinct steady-state solutions, share similarly shaped solution branches, and rely on control parameters with comparable physical interpretations. These parallels suggest concrete hypotheses about the underlying mechanisms of biological switching. 
In particular, the hydraulic analogy highlights the central role of energy flux, driven in cells by imbalances in ATP/ADP or GTP/GDP concentrations, in initiating and sustaining switch activation. It also points to two parameters that predominantly govern the transition between on and off states: the energy difference separating these states and the effective height of the energy barrier along the transition pathway. Together, these insights indicate that biological signaling dynamics may be governed by general physical principles that extend beyond the biochemical details of any single receptor.

\begin{figure}[h]
\includegraphics[width=10.5 cm]{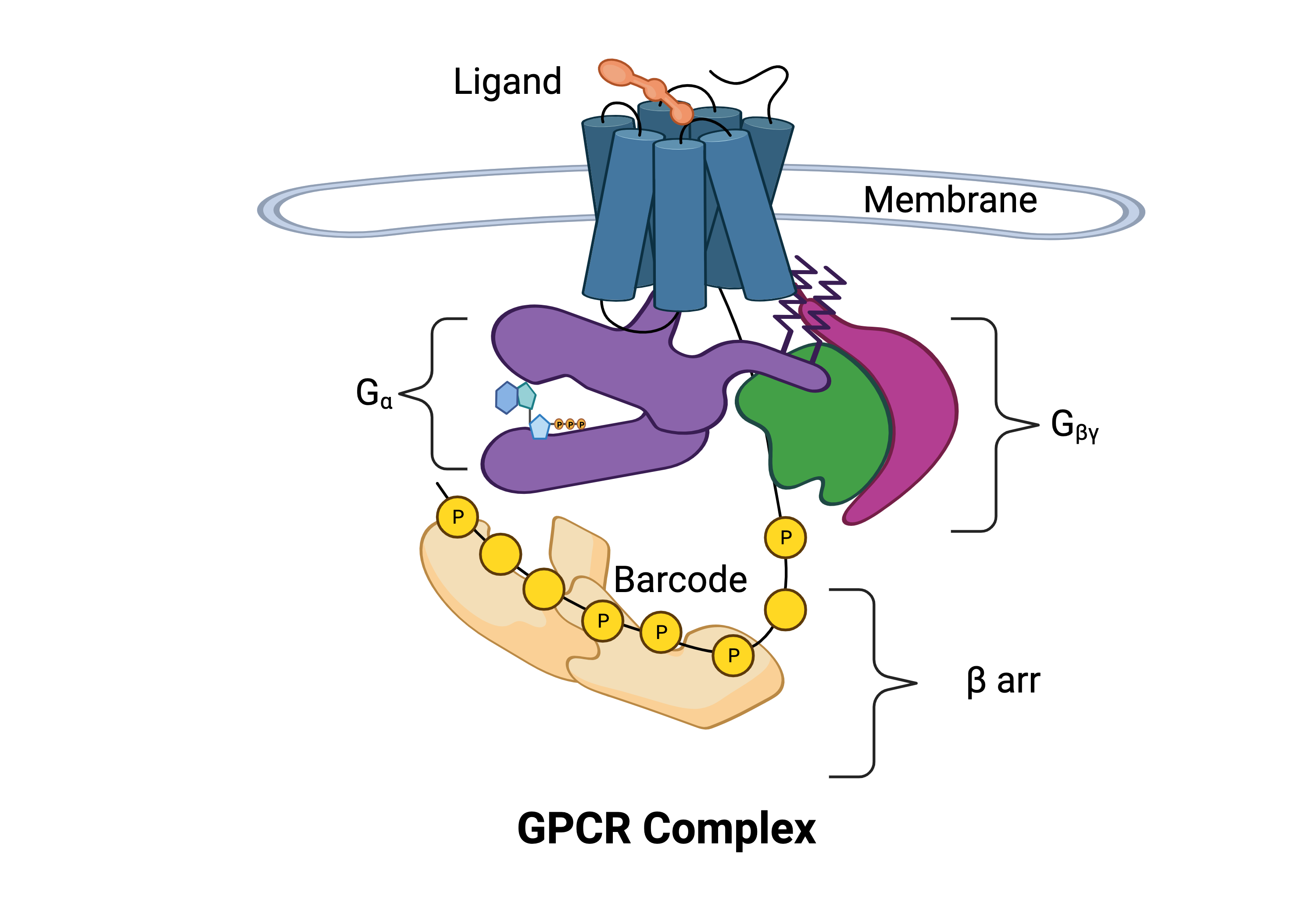}
\caption{The GPCR structure is shown with seven transmembrane $\alpha$-helices (blue) forming the receptor core. Ligand binding in the extracellular pocket (orange) induces conformational changes that propagate through the helices to the intracellular region, which is separated here for clarity. Two signaling pathways can be engaged: the G$\alpha$-mediated pathway (purple), involving the G$\alpha$ subunit of the heterotrimeric G protein (with $\beta$ and $\gamma$ subunits), and the $\beta$-arrestin pathway (tan), directed by phosphorylation patterns on the receptor’s C-terminal tail and intracellular loops that form a ligand-specific barcode.
}
\label{fig1}
\end{figure}

Perhaps the most important biological systems illuminated by this study are G protein–coupled receptors (GPCRs), shown schematically in Fig. \ref{fig1}.
Although GPCR papers often begin by noting that these receptors constitute the largest family in the human genome and account for roughly 30\% of drug targets \cite{santos2017comprehensive}, the broader scientific implications of this class of proteins extend beyond their pharmaceutical relevance. 
GPCRs are prototypical molecular information processors whose behavior reflects the general principles of nonequilibrium computation.

The divergent signaling outputs generated through the G-protein and $\beta$-arrestin coupling illustrate how a single receptor architecture can transform small extracellular perturbations into high-dimensional intracellular codes. Opioid receptors provide a particularly clear demonstration of this divergence: G-protein engagement supports analgesia, while $\beta$-arrestin recruitment contributes to tolerance and addiction. This duality underscores the need to understand GPCR signaling not merely in terms of ligand affinity or efficacy but also as a problem of information routing through competing molecular pathways.

The structural organization of the GPCR complex offers a mechanistic explanation for these behaviors. Ligand binding induces coordinated rearrangements of transmembrane helices, modulating two classes of molecular switches: the conformational switch G$\alpha$ and a distributed set of phosphorylation sites undergoing phosphorylation–dephosphorylation cycles. The resulting phosphorylation barcode encodes ligand-specific information that determines the engagement of $\beta$ -arrestin and downstream signaling outcomes. This multilayered architecture demonstrates how a low-dimensional extracellular signal is systematically transformed into a richer intracellular representation, consistent with the view of GPCRs as molecular nanoprocessors.

Together, the findings of this study support a conceptual shift from viewing GPCRs as simple ligand-activated receptors to treating them as nonequilibrium information-processing machines. This perspective links molecular pharmacology to broader theoretical frameworks in information theory and statistical physics, providing a basis for interpreting biased signaling, ligand-specific barcodes, and pathway selection as emergent computational strategies. Such an approach may open new avenues in therapeutic design by emphasizing control over information flow rather than isolated biochemical interactions.

The macroscopic plumbing analogy developed in this study serves as an intuitive representation of these principles. By capturing the essential elements of driving flux, energy barriers, and stochastic fluctuations, the hydraulic model clarifies otherwise opaque aspects of GPCR switching behavior. This simplified framework also generalizes naturally to interacting networks of switches, suggesting that the computational capacity of GPCRs arises not only from individual molecular mechanisms but from their integration within larger signaling architectures.

\section{Conclusions}

\begin{itemize}
    \item 
    We identified a macroscopic hydraulic system that is an analog of the mesoscopic molecular GPCR information-processing systems found on the cell membrane of animals.
    \item 
    The analog has a structure and functionality similar to the mesoscopic original, including
    \begin{itemize}
        \item
         both models contain up to three solutions; 
         \item the overall form of the solutions is similar
        \item 
        the physical meaning of the control parameters is similar;
    \end{itemize}
    \item 
    The similarities between the analog and the original systems suggest hypotheses to be tested on the original system
    \begin{itemize}
        \item 
        The energy flux through the system driven by the imbalance in concentrations of ATP/ADP or GTP/GDP is primarily responsible for the activation of the switch.
    \item 
    The key parameters that primarily control the switch on/off state are 
    \begin{itemize}
        \item 
        The energy difference between the on and off states,
        and
        \item 
        The height of the energy barrier in the path from the on state to the off state.
    \end{itemize}
\end{itemize}    
\end{itemize}

\subsubsection{Acknowledgements}
  Research by AG is supported by MIUR PRIN-COFIN2022  grant 2022JWAF7Y.

\begin{credits}

\subsubsection{\discintname}
The authors have no competing interests to declare
that are relevant to the content of this article. 
\end{credits}

\bibliographystyle{splncs04}
\bibliography{library}

\end{document}